\documentclass[runningheads]{llncs}

\usepackage[utf8]{inputenc} 
\usepackage[T1]{fontenc}    
\usepackage{hyperref}       
\usepackage{url}            
\usepackage{booktabs}       
\usepackage{graphicx}
\usepackage{doi}

\usepackage{amsmath}
\usepackage{bbding}
\usepackage{amssymb}
\usepackage{subcaption}

\usepackage{hhline}
\usepackage{cleveref}
\usepackage{xcolor}
\usepackage{multirow}
\usepackage{breqn}
\usepackage{tabu}

\newcommand{\repeatthanks}{\textsuperscript{\thefootnote}}
\newcommand{\namemethod}{Y-Net}
\title{\namemethod: A Spatiospectral Dual-Encoder Network for Medical Image Segmentation}

\titlerunning{\namemethod{}: A Spatiospectral Network for Medical Image Segmentation}

\date{} 					

\author{Azade Farshad\thanks{Equal Contribution}\inst{1} \and
Yousef Yeganeh\repeatthanks \inst{1} \and
Peter Gehlbach\inst{2} \and
Nassir Navab\inst{1,2}
}
\authorrunning{A. Farshad et al.}
\institute{Technical University of Munich, Munich, Germany \and
Johns Hopkins University, Baltimore, USA
}

\begin{document}
\maketitle

\begin{abstract}
Automated segmentation of retinal optical coherence tomography (OCT) images has become an important recent direction in machine learning for medical applications. We hypothesize that the anatomic structure of layers and their high-frequency variation in OCT images make retinal OCT a fitting choice for extracting spectral domain features and combining them with spatial domain features. In this work, we present \namemethod, an architecture that combines the frequency domain features with the image domain to improve the segmentation performance of OCT images. The results of this work demonstrate that the introduction of two branches, one for spectral and one for spatial domain features, brings very significant improvement in fluid segmentation performance and allows outperformance as compared to the well-known U-Net model. Our improvement was $13\%$ on the fluid segmentation dice score and $1.9\%$ on the average dice score. Finally, removing selected frequency ranges in the spectral domain demonstrates the impact of these features on the fluid segmentation outperformance. Code: \url{github.com/azadef/ynet}
\end{abstract}
\section{Introduction}
Ocular Optical Coherence Tomography (OCT) is among the heavily utilized clinical imaging modalities by ophthalmologists and retina specialists. Segmentation of OCT images now drives the diagnosis and treatment of eye diseases such as diabetic macular edema \cite{virgili2015optical} (DME) and age-related macular degeneration (AMD). Segmentation of intraretinal fluid pockets is especially useful as it determines the presence, extent and response of retina to treatment. However, despite the importance of fluid segmentation in OCT images, existing methods fail to segment this area efficiently. In this work, we propose to extract and process information from the spectral domain due to the existence of spectral features in OCT images, that may otherwise be missed in existing spatial neural networks. Furthermore, it has been shown in previous work \cite{chi2020fast} that spatial information only focuses on local information and fails to target the global information across all the pixels in an image. This problem is solved here by combining features from both spectral and spatial domains.\\
To summarize our contributions: 1) We propose \namemethod, an end-to-end autoencoder based architecture, with two encoder branches for automated retinal layer and
fluid segmentation, in OCT images. 2) Our proposed spectral encoder is designed for extracting frequency domain features from the images. 3) \namemethod{} outperforms the well-known U-Net \cite{Ronneberger15UNet} architecture and other related work by a minimum of $13\%$ in fluid segmentation, and $1.9\%$ on average in terms of dice score. 4) The \namemethod{} architecture has less parameters compared to U-Net.

\section{Related Work}
Many of the early methods for segmentation of retinal OCT images \cite{chiu10automatic} relied on graph-based techniques (e.g., graph cut, shortest path). Subsequently, some works focused on the combination of neural networks and graph-based methods for estimating the final retinal layer boundaries \cite{fang17} or combining graph convolutional networks with other neural networks \cite{li2021multi}.

He et al. studied OCT segmentation in a series of works \cite{He19,he2018topology,he2019fully} considering OCT scan topology. Utilizing fully convolutional networks (FCN) has been explored \cite{he17towards,Kugelman19} for predicting segmentation maps and correcting the topology based on a specific topology criterion.

A number of recent methods in medical image segmentation focus on using autoencoder based deep neural networks \cite{roy2017relaynet,Kiaee18} for end-to-end segmentation. One of the earliest and best-known autoencoder based architectures for 2D medical image segmentation is U-Net \cite{Ronneberger15UNet}. The evolution of U-shaped networks for image segmentation has been of high research interest in recent years. Many works, such as MDAN-U-Net \cite{Liu20mdan} try to use multiscale features or an attention mechanism to improve the segmentation performance of existing methods. Feature Pyramid Networks (FPNs), which are commonly used in the computer vision community, have also been of interest in medical image segmentation for global feature extraction \cite{feng2020cpfnet,li2020deepretina}. Other lines of work focus on networks designed specifically for the OCT segmentation task \cite{reddy2020retinal,wei20,Kiaee18}, using Gaussian process \cite{pekala2019deep}, feature alignment \cite{Duan18,maier2021line}, or epistemic uncertainty \cite{orlando2019u2}.

Using Recurrent Neural Networks (RNNs) for OCT segmentation has been explored in \cite{kugelman18,tran2020retinal}. While \cite{kugelman18} considered sequences between different scans, Tran et al. \cite{tran2020retinal} modeled OCT retinal layers using natural language and developed an OCT segmentation method using RNNs for processing pixel sequences.
An autoencoder network with two encoder branches has been previously used for polyp detection \cite{mohammed2018net} by taking advantage of a pre-trained VGG network. The purpose of this application is considerably different than ours. A combination of U-Net  \cite{Ronneberger15UNet} and fast Fourier transforms (FFT) \cite{Nair2020234} has been explored for reducing the computational costs of convolutional networks. Recently fast Fourier convolutions \cite{chi2020fast} were integrated into the image inpainting task \cite{suvorov2022resolution} by the computer vision community. The purpose was to use the global patterns that exist in images which may not be well extracted by regular convolutional layers. This inspired us to take advantage of the fast Fourier convolutions for the task of OCT segmentation due to the existence of high-frequency speckles, which are a function of the tissue and its layers \cite{schmitt1999speckle}. The existence of these speckles can harm the model performance when using only spatial features; Therefore, we hypothesize that by extracting spectral features from the OCT images, our network will be able to disentangle features from different frequency distributions. This enables the model to attend to more important frequency ranges in the features using adaptive learnable kernels in FFT Convolutions and to be able to model the high-frequency variation and distribution within each layer.
\section{Method}
In this section, we present the core principles of our work. First, we explain the overall structure of the segmentation framework; we then describe the components of our proposed spectral encoder and, at its core, the Fourier unit that performs the spectral feature extraction function. Finally, the loss functions used in this work are presented.
\subsection{Segmentation Framework}
\begin{figure}[tb]
    \centering
    \includegraphics[width=\textwidth]{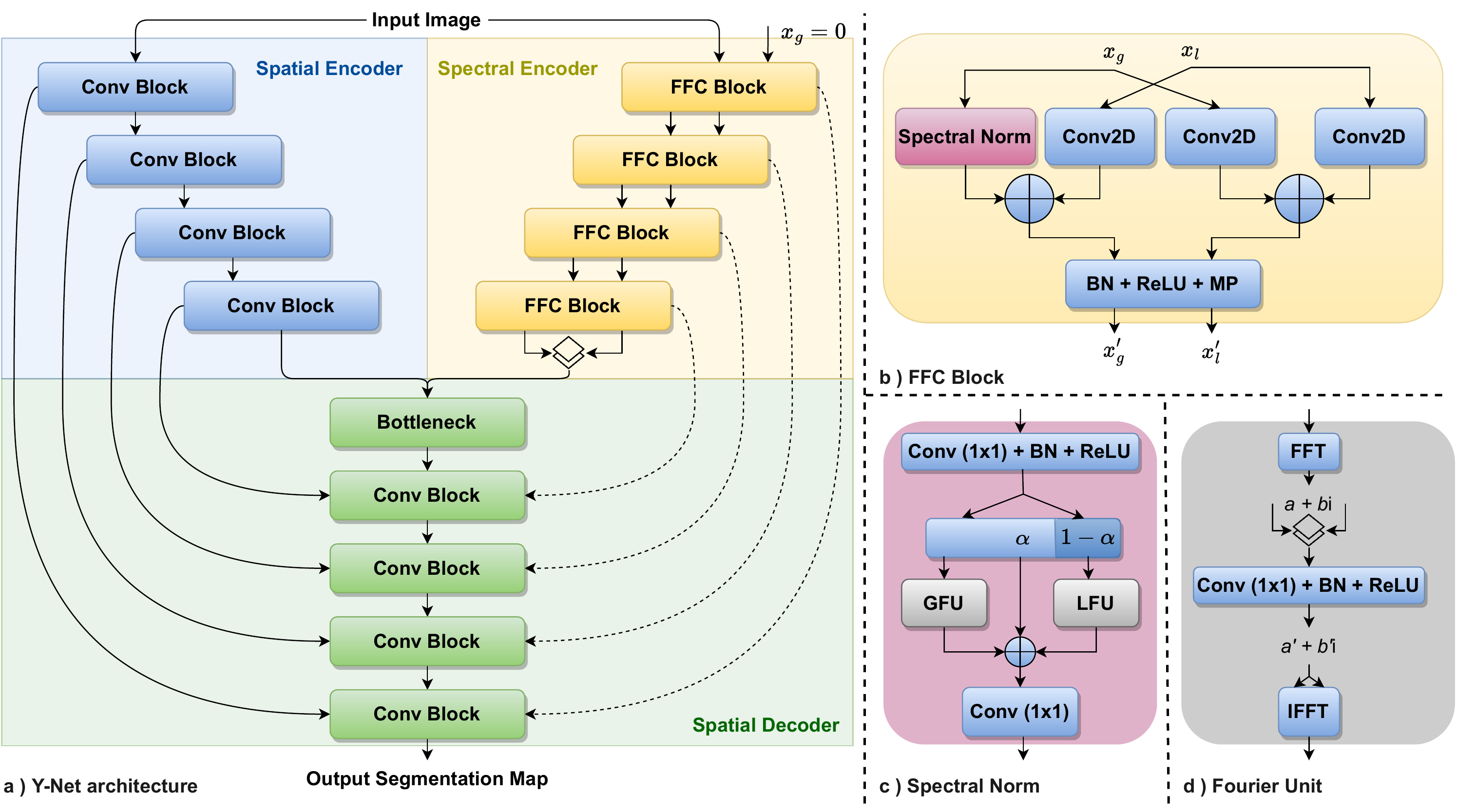}
    \caption{\textbf{a) \namemethod:} Our proposed network has two branches, one for processing spatial features similar to previous works and our proposed branch for extracting spectral features. The spectral encoder has four FFC blocks which gets the local and global features $x_l$, $x_g$ as input and generates the processed features $x'_l$, $x'_g$. \textbf{b) FFC Block:} The FFC blocks process the local features using Conv2D layers and process the global features using the spectral norm. \textbf{c) Spectral Norm:} The global information is divided into two portions which are fed to a Fourier unit. \textbf{d) Fourier Unit:} Here, the fast Fourier transform, followed by a conv layer, is applied to the features to get the frequency domain features. Finally, the processed features are brought back to the spatial domain using inverse FFT.}
    \label{fig:method}
\end{figure}

The segmentation network predicts the segmentation map $\hat{y}$ given an input image  $x \in \mathbb{R}^{H \times W}$ and its corresponding segmentation label $y \in \mathbb{Z}^{H \times W}$, where $H$ and $W$ are the image height and width, respectively. As shown in \autoref{fig:method}-a, the segmentation network, \namemethod{} consists of two encoder branches $E_c$, $E_f$, where $E_c$ is the spatial encoder with convolutional blocks, and $E_f$ is our proposed spectral encoder with fast Fourier convolutional (FFC) blocks \cite{chi2020fast}. The decoder network $G(.)$, receives the combined spatial and spectral features from the encoder networks and generates the segmentation map $\hat{y}$, where $\hat{y} = G(E_c(x), E_f(x))$. Similar to U-Net \cite{Ronneberger15UNet}, \namemethod{} has an autoencoder based structure with skip connections from spatial encoder blocks to decoder blocks. The role of the proposed spectral encoder is to extract and process global features from the frequency domain that may have been missed by the spatial convolutions. This section explains each of our network's components and the objective functions in detail.

\subsubsection{Spatial Encoder}
The spatial encoder in our network is the same as the original U-Net \cite{Ronneberger15UNet} with four convolutional blocks. Each convolutional block consists of a convolutional layer, batch normalization layer (BN), an activation function (ReLU) and a max pooling (MP) layer. The input to the first convolutional block is the input image, and the output of each block is fed to the next block as shown in \autoref{fig:method}-a.
\subsubsection{Spectral Encoder}
Here we introduce our spectral encoder, to extract spectral domain features from the data. Our spectral encoder receives the same input as the spatial encoder. The input image $x$ is fed to the first FFC block as local information $x_l$. The value of $x_g$ is set to zeros for the first block since the input image pixels are considered local information, and there are no global features in the input image. Similar to the spatial encoder, there are in total four FFC blocks in the spectral encoder.
\subsubsection{Spatial Decoder}
The spatial decoder network $G$ has four up convolutional blocks in total. It receives the spectral and spatial features and concatenates them before passing them to the bottleneck layer. Then, the features from the previous decoder block and features from the skip connections are up-scaled using a convolutional block similar to the spatial encoder, followed by transpose convolutional layers. The final segmentation map is generated by the final decoder block.
\subsection{Spectral Encoder Components}
\subsubsection{Fast Fourier Convolutional Block}
The Fast Fourier convolutional (FFC) block, shown in \autoref{fig:method}-b receives the global and local information $x_g$, $x_l$ as input. Then $x_g$ and $x_l$ are fed to three convolutional layers to extract the global and local spatial features and the spectral norm, which performs the frequency domain feature extraction. Finally, batch normalization, a non-linear activation function, and max pooling are applied to the features to generate the global and local features, $x'_g$, $x'_l$ for the next FFC block.
\subsubsection{Spectral Norm}
The spectral norm (\autoref{fig:method}-c) first applies a convolutional block with a kernel size of 1 to $x_g$, which produces $x''$. The channels of $x''$ are then divided into two portions based on a predefined value $\alpha$, with $\alpha$ percentage of channels considered the global information and $1-\alpha$ percentage of channels the local information. The divided global and local features are separately fed to Fourier units ($FU_g$, $FU_l$) to extract spectral features $x''_g$, $x''_l$. It should be noted that $FU_g$ and $FU_l$ share the same architectural design. Finally, $x''$ and the output of global and local Fourier units $x''_g$, $x''_l$ are summed and fed to a convolutional layer with kernel size 1.
\subsubsection{Fourier Unit}
The Fourier unit (\autoref{fig:method}-d) receives a portion of $x''$ as input, then fourier transform is applied to those features to obtain real and imaginary parts $a+bi \in \mathbb{C}$. The real and imaginary parts $a,b$ are stacked and then fed to a convolutional layer with a kernel size of 1. An activation layer and a batch normalization layer are applied to the output of the convolutional layer. The output is then split into two parts, namely the real and imaginary parts $a'$, $b'$, which are then fed to the inverse Fourier transform to convert the features back to the spatial domain. 

\subsection{Losses}
Our models are trained with a combination \cite{taghanaki2019combo} of dice loss \cite{sudre2017generalised} and cross-entropy loss. The combo loss is widely used for medical image segmentation . The loss between each ground truth segmentation label $y$ and the predicted segmentation map $\hat{y}$ is computed as follows:
\begin{equation} \label{eq:dice}
    \mathcal{L}_{Dice}(y,\hat{y}) = 1- \frac{2y\hat{y}+\epsilon}{y+\hat{y}+\epsilon}
\end{equation}
The dice Loss considers the intersection over union (IoU) and is computed as shown in \autoref{eq:dice}. To ensure the numerical stability, a very small value, $\epsilon$ is used for computing the dice loss. The cross-entropy loss as shown in \autoref{eq:ce} maximizes the cross-entropy information between the true and predicted labels.
\begin{equation}\label{eq:ce}
    \mathcal{L}_{CE}(y,\hat{y}) = - \frac{1}{N} \sum_{i=0}^N y_{i} \log (\hat{y}_{i})
\end{equation}

With $\lambda_{Dice}$, $\lambda_{CE}$ being the weighting factor for each loss term, the total loss then becomes:

\begin{equation}\label{eq:total}
    \mathcal{L}_{total} = \lambda_{Dice} \mathcal{L}_{Dice} + \lambda_{CE} \mathcal{L}_{CE}
\end{equation}
\section{Experiments}
\begin{table}[tb]
     \centering
     \caption{Mean and per layer dice score compared to related works on the publicly available Duke OCT dataset \cite{chiu15duke}}
     \resizebox{\textwidth}{!}{
     \begin{tabular}{|c|c|c|c|c|c|c|c|c|c|}
     \hline
        Method  & ILM & NFL-IPL & INL & OPL & ONL-ISM & ISE & OS-RPE & Fluid & Mean \\ \hline
        RelayNet \cite{roy2017relaynet} & 0.84 & 0.85 & 0.70 & 0.71 & 0.87 & 0.88 & 0.84 & 0.30 & 0.75  \\ \hline
        Language \cite{tran2020retinal} & 0.85 & 0.89 & 0.75 &  0.75 & 0.89 & \textbf{0.90} & \textbf{0.87} & 0.39 & 0.78  \\ \hline 
        Alignment \cite{maier2021line} & 0.85 & 0.89 & 0.75 & 0.74 & 0.90 & \textbf{0.90} & \textbf{0.87} & 0.56 & 0.81 \\\hline
        U-Net \cite{Ronneberger15UNet} & 0.84 & 0.89 & 0.77 & \textbf{0.76} & 0.89 & 0.89 & 0.85 & 0.80 & 0.836 \\\hline 
        \namemethod{} (Ours) & \textbf{0.86} & \textbf{0.89} & \textbf{0.78} & 0.75 & \textbf{0.90} & 0.88 & 0.85 & \textbf{0.93} & \textbf{0.855} \\\hline
     \end{tabular} }
     \label{tab:resuls}
\end{table}

In this section, we evaluate our proposed method and compare it to the existing literature and known baselines. First, we discuss our experimental setup, and then present a comparison of our model to prior work, and finally, we show an ablation study of the components in our model. As discussed in previous sections, we tackle the problem of retinal layer and fluid segmentation using OCT images. We train and test our proposed method on the publicly available Duke OCT dataset \cite{chiu15duke} and compare it against multiple previous works on OCT segmentation. All reported results except U-Net \cite{Ronneberger15UNet} are taken from the original values reported in the papers. The results of RelayNet \cite{roy2017relaynet} are taken from \cite{tran2020retinal} based on the 6-2-2 split for the evaluation. Please refer to the supplementary material for the results on the UMN dataset \cite{rashno2017fully} and based on the mIoU metric on both datasets. 
\begin{figure}[htb]
    \centering
\includegraphics[width=0.8\textwidth]{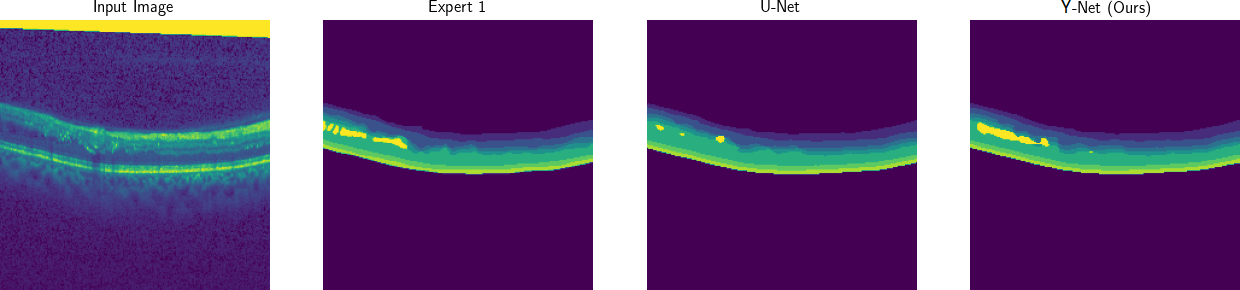}
\includegraphics[width=0.8\textwidth]{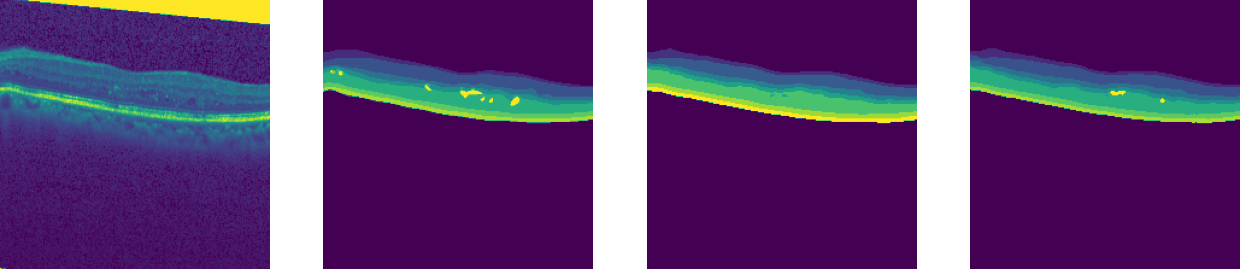}
\includegraphics[width=0.8\textwidth]{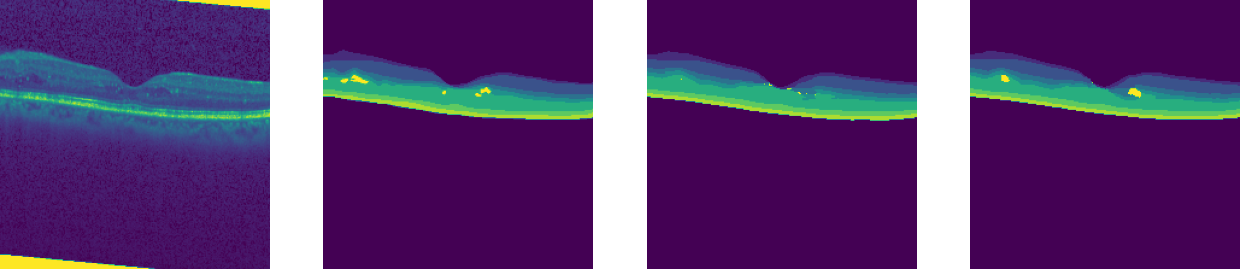}
\caption{Some qualitative results of \namemethod{} compared to U-Net \cite{Ronneberger15UNet}} \label{fig:qualitative}
\end{figure}
\subsection{Experimental Setup}
We follow the same experimental protocol for training and evaluation of our method on the Duke OCT dataset, as in prior works  \cite{tran2020retinal,maier2021line}. The Duke OCT dataset consists of OCT scans from 10 patients, which are annotated by two experts. The scans from the first six subjects are used for training, subjects 7 and 8 for validation, and the scans from the remaining two subjects are used for testing. Our models are trained and tested on the annotations from expert 1, similar to previous works.
All our models and the U-Net \cite{Ronneberger15UNet} were trained with a batch size of $10$, a learning rate of $5e-4$, weight decay of $1e-4$, maximum $80$ epochs of training and Adam optimizer. The number of training epochs was chosen based on the best validation accuracy for all models. The values of $\lambda_{Dice}, \lambda_{CE}$ were found empirically and set to $1$ for both. 
The images were resized to $(224 \times 224)$. The evaluations are reported using dice score values for all retinal layers, fluid and their average. The number of model parameters for U-Net and \namemethod{} are 7.76M, 7.46M respectively.
\subsection{Results}
\autoref{tab:resuls} shows the dice score of our proposed method for various retinal layers and their fluid pockets, as compared to previously reported approaches.  Our model tests on par or outperforms prior work for segmentation of retinal layers, and has a large gap with other models in fluid segmentation performance. We argue that this performance gain is due to existence of features in certain frequency ranges that relate to the fluid pockets. We verify this hypothesis in one of our experiments by modifying the range of of frequency values in the Fourier unit. We also show some qualitative results to compare the fluid segmentation performance of our model to U-Net in \autoref{fig:qualitative}. We report that our model is able to segment the fluid pockets similar to expert one's annotation, while U-Net fails to segment the fluid in some regions.

\subsubsection{Ablation Study}
In \autoref{tab:resuls_ablation}, we present an ablation study of the components in our model. The first row shows the performance of the \namemethod{} architecture with regular convolutional blocks in the second branch. We evaluate our model on this architecture to show that the current improvement in average dice score, and especially the fluid segmentation performance is not from merely increasing the size of the network, and is affected by the introduction of the FFC blocks. The rest of the table shows the performance of our model given different values of $\alpha$. As discussed in the methodology, $\alpha$ defines the percentage of features in the global and local Fourier units. As it can be seen in \autoref{tab:resuls_ablation}, the best performance is gained by the $\alpha$ values of $0.25$ and $0.5$ with both models reaching a dice score of $0.93$ in fluid segmentation while being on par with other models in retinal layer segmentation. We argue that both local and global features have valuable information that our model could learn. By having a value of $\alpha$ which is neither too large (1) nor too small (0), our model is able to correlate the global and local features to achieve the best performance.

\begin{table}[thb]
     \centering
     \caption{Ablation study on the FFC blocks and the value of $\alpha$}
     \resizebox{\textwidth}{!}{
     \begin{tabular}{|c|c|c|c|c|c|c|c|c|c|c|c|}
     \hline
        FFC Block & $\alpha$ & ILM & NFL-IPL & INL & OPL & ONL-ISM & ISE & OS-RPE & Fluid & Mean \\ \hline
        - & - & \textbf{0.87} & \textbf{0.90} & 0.77 & 0.75 & 0.89 & 0.88 & \textbf{0.86} & 0.89 & 0.851 \\ \hline
        \Checkmark & 0 & 0.86 & 0.89 & 0.76 & 0.75 & \textbf{0.90} & \textbf{0.89} & 0.85 & 0.86 & 0.845 \\ \hline
        \Checkmark & 0.25 & 0.86 & 0.89 & 0.77 & 0.74 & 0.89 & \textbf{0.89} & \textbf{0.86} & \textbf{0.93} & 0.854 \\ \hline
        \Checkmark & 0.5 & 0.86 & 0.89 & \textbf{0.78} & 0.75 & \textbf{0.90} & 0.88 & 0.85 & \textbf{0.93} & \textbf{0.855} \\ \hline
        \Checkmark & 0.75 & 0.84 & 0.87 & 0.76  & 0.73 & 0.89 & 0.88 & \textbf{0.86} & 0.90 & 0.841 \\ \hline
        \Checkmark & 1 &  0.85 & 0.89 & 0.77 & \textbf{0.76} & 0.89 & 0.88 & 0.85 & 0.88 & 0.846 \\ \hline
     \end{tabular} }
     \label{tab:resuls_ablation}
\end{table}
\begin{table}[htb]
     \centering
     \caption{Effect of variation in frequency ranges}
     \resizebox{\textwidth}{!}{
     \begin{tabular}{|c|c|c|c|c|c|c|c|c|c|c|c|c|}
     \hline
        Spectral features range & ILM & NFL-IPL & INL & OPL & ONL-ISM & ISE & OS-RPE & Fluid & Mean \\ \hline
        No change & 0.85 & \textbf{0.89} & 0.77 & \textbf{0.75} & 0.90 & \textbf{0.89} & 0.85 & 0.93 & 0.854 \\ \hline
        keep(-10,10) & \textbf{0.86} & \textbf{0.89} & \textbf{0.78} & \textbf{0.75} & 0.90 & 0.88 & 0.85 & \textbf{0.93} & \textbf{0.855} \\ \hline
        remove(-10,10) & 0.84 & 0.88 & 0.76 & 0.74 & 0.90 & 0.88 & 0.85 & 0.78 & 0.829 \\ \hline
     \end{tabular} }
     \label{tab:resuls_frequency}
\end{table}
We further explore the effect of the FFC blocks in \autoref{tab:resuls_frequency} by varying the range of frequencies processed by the Fourier units in specific ranges. The first row in \autoref{tab:resuls_frequency} shows the regular model without any changes to the frequency range, which is usually between $-40$ and $40$. In the second row, we clip the range of the frequency values to $(-10,10)$, which slightly increases the overall segmentation performance. In the last row, we remove the frequencies between $(-10,10)$ by setting the value of the frequencies from $-10$ to $0$ to $-10$ and setting the positive values between $0$ and $10$ to $10$. As it can be seen, the fluid segmentation performance drops to $0.78$ in this setting, while the retinal layer segmentation performance drops very slightly. Based on these experiments, we can argue that the high segmentation performance of the fluid pockets by our model is affected by the spectral domain features and that the features used for the fluid segmentation belong to a specific range of frequencies (here close to $(-10,10)$).

In our experiments, we also tried using focal frequency loss \cite{jiang2021focal} and adding skip connection from the spectral encoder to the spatial decoder, but this did not improve model performance. We believe that adding skip connections from the spectral domain to the spatial domain does not convey significant advantage since the spectral features and global information may not correlate well with the segmentation map.
\subsection{Discussions and Conclusion}
In this work, we present an end-to-end autoencoder based architecture for the segmentation of retinal layers and fluid pockets in ocular OCT images. Our proposed network \namemethod{} extracts spectral domain features in a second encoder branch proposed by us in addition to the spatial encoder used in previous works. We hypothesized that by extracting spectral domain features from OCT images, that have high-frequency non-uniform speckles that are dependent on the tissue and retinal layers, our model would learn features that would improve OCT segmentation performance. Learning features in the frequency domain would enable our network to model and learn the distribution of speckles within each layer. Our experiments showed that the model would be highly affected by varying the range of frequencies in the Fourier units in the core of our FFC blocks. This could verify our hypothesis that certain frequencies in the OCT images may correlate with specific layers or fluid pockets. We compared our final proposed model to multiple previous works and showed that our model outperforms existing models by $13\%$ in fluid segmentation, reaching a value of $0.93$ in dice score, while achieving on par or better performance in retinal layer segmentation.

\section*{Acknowledgement}
We gratefully acknowledge the Munich Center for Machine Learning (MCML) with funding from the Bundesministerium f\"ur Bildung und Forschung (BMBF) under the project 01IS18036B.

\bibliographystyle{splncs04} 
\bibliography{ref}

\end{document}